**Broadband photon pair generation from a single lithium niobate microcube**

*Ngoc My Hanh Duong,[1]\* Gregoire Saerens,[1] Flavia Timpu,[1] Maria Teresa Buscaglia,[2] Vincenzo Buscaglia,[2] Andrea Morandi,[1] Jolanda S. Muller,[1] Andreas Maeder,[1] Fabian Kaufmann,[1] Alexander Sonltsev,[3] and Rachel Grange[1]\**

[1]ETH Zurich, Optical Nanomaterial Group, Institute for Quantum Electronics, Department of Physics, 8093 Zurich, Switzerland

[2]Institute of Condensed Matter Chemistry and Technologies for Energy, National Research Council, via De Marini 6, 16149 Genoa, Italy

[3]School of Mathematical and Physical Sciences, University of Technology Sydney, Ultimo, NSW, 2007, Australia

\*Corresponding author: nduoeng@phys.ethz.ch, grange@phys.ethz.ch

Keywords: lithium niobate, spontaneous parametric down-conversion, quantum optics, nonlinear optics, photon pairs

Nonclassical light sources are highly sought after as they are an integral part of quantum communication and quantum computation devices. Typical sources rely on bulk crystals that are not compact and have limited bandwidth due to phase-matching conditions. In this work, we demonstrate the generation of photon pairs from a free-standing lithium niobate microcube at the telecommunication wavelength through the spontaneous parametric down-conversion process. The maximum photon pair generation rate obtained from a single microcube with the size of ~4 μm is ~80 Hz, resulting in an efficiency of ~1.2 GHz/Wm per unit volume, which is an order of magnitude higher than the efficiency of photon-pair generation in bulky nonlinear crystals. The microcubes are synthesized through a solvothermal method, offering the possibility for scalable devices via bottom-up assembly. Our work constitutes an important step forward in the realization of compact nonclassical light sources with broadband tunability for various applications in quantum communication, quantum computing, and quantum metrology.

## 1. Introduction

A photon-pair source is an important building block for applications in quantum communication, quantum networks, and quantum computation [1][2][3][4]. So far, spontaneous wave-mixing processes, including four wave-mixing [5][6] and three-wave mixing have been considered versatile techniques for the generation of nonlinear correlated photon pairs [1]. Between the two techniques, a three-wave mixing process of spontaneous parametric down-conversion (SPDC), which relies on second-order nonlinearity, can achieve higher conversion efficiency compared to its third-order counterpart in the same nonlinear material volume, except for the cases of nontrivial mode overlap [7]. SPDC is considered the most promising source for the generation of photon pairs owing to its robust operation at room temperature. Furthermore, the generated photon pairs can be entangled in polarization [8][9], angular momentum [10] and frequency [11], as well as spatially [12] and temporally coherent [13].

Alternative platforms for nonclassical light emitters have been realized in solid-state optically active defects including fluorescent color centers [14][15], quantum dots (QDs) [16], and two-dimensional materials such as transition metal dichalcogenides (TMDCs) and hexagonal boron nitride [17][18][19]. These systems can be used to generate single photons, competing with heralded single-photon generation via spontaneous parametric nonlinear processes [20], and can also be utilized for photon-pair generation, including entangled photon pairs [21]. However, QDs and the TMDCs only exhibit nonclassical photon emission at cryogenic temperatures, which involves expensive and bulky cooling systems. All these sources also suffer from inhomogeneous distributions and spectral diffusion due to the charges in the solid-state host environment [22], thus making it difficult to generate correlated two photons from different emitters. Moreover, interfacing those atomic defects to photonic structures remains a non-trivial task due to the high refractive index of host materials, which makes extraction of photons challenging. Apart from that, there is only a limited number of quantum light sources that can generate single photons at telecom wavelengths [23], which are used in conventional long-distance optical communication systems where the losses are the lowest. Therefore, the quest for an ideal photon-pair source is still continuing.

The efficiency of the SPDC process is improved for systems with satisfied phase-matching conditions. Conventionally, phase-matching conditions can be achieved by temperature and wavelength tuning of centimeter-sized crystals [9][24][25]. However, this method offers relatively low photon-pair rates while requiring a rather complicated setup with multiple optical components. To reduce the footprint of the SPDC devices, photonic architectures such as waveguides [26] and resonators are sculpted out of nonlinear materials, which can significantly improve the conversion efficiency while reducing the size of the devices to millimeters and tens of micrometers [20]. These approaches, however, also result in lengthy devices with scales of at least tens and typically hundreds of microns as the longitudinal phase-matching conditions still need to be satisfied [27].

A promising emerging platform is miniaturized SPDC quantum light sources. Following the demonstration of SPDC in thin films [28][29], it has been shown that nanoresonators supported by Mie resonances can be versatile sources for quantum light generation through nonlinear processes owing to the high refractive index and strong intrinsic nonlinearity [30]. Initial experiments on nonclassical light generation have been realized in AlGaAs nanoresonators [31] as well as LiNbO$_3$ metasurfaces [32]. Nevertheless, these approaches still require careful modal matching [33] and sophisticated fabrication techniques, which stand as significant scale-up challenges. Also, the nonlinear emission from the substrate can further complicate the two-photon generation process from the nanoresonators, making it difficult to capture and route the emitted photons from these structures with high efficiency. This research direction remains largely unexplored experimentally, hence, it is desirable to study this phenomenon at the micro- and nano-scale regime while solving the substrate and phase matching issues.

Here, we present a high photon pair generation rate of ~80 Hz at the telecommunication wavelengths from single bottom-up grown lithium niobate (LN) microcubes with sizes ranging from 2 to 4 μm. The bottom-up growth method warrants the device scalability and cost effectiveness. The rationale behind the choice of LN is due to its optical properties such as a wide transparent window ranging from 0.4 - 5.0 μm, a low absorption coefficient, and especially a high second-order ($\chi^{(2)}$) optical susceptibility component ($d_{33}$ = ~34 pm/V). The generation of photon-pairs through SPDC from LN microcubes opens the opportunity for integrable two-photon light sources down to microscale, which allows the miniaturization of functional quantum devices. Compared to QDs or defect-based photon sources, SPDC occurs at room temperature, eliminating the need for a cryogenic chamber, and thus reducing the cost for equipment. Compared to other nonlinear approaches, owing to the relaxation of longitudinal phase-matching, our system can produce photon pairs in a wide range of wavelengths [34][32], enabling broadband tunability for nonlinear quantum light sources.

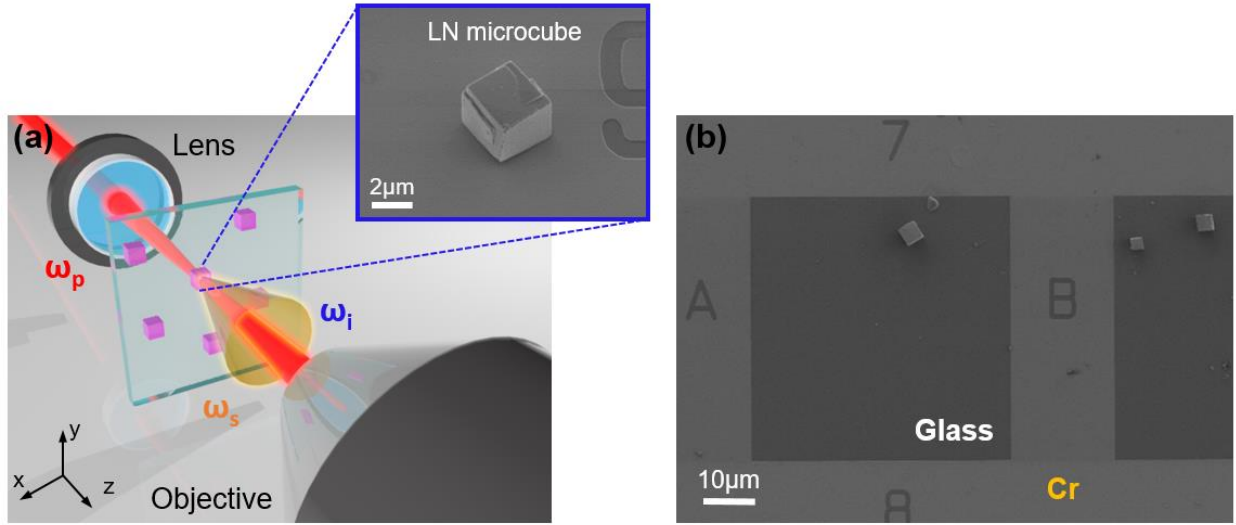

**Figure 1**. (a) Schematics of the transmission setup for LN microcube optical characterization. Top image: SEM image of an example of a typical microcube in this study. (b) SEM image of the sample showing a distribution of various LN microcubes on the substrate with markers. The brighter area around the 50 µm dark square area is covered with Cr to enable electronic transport during SEM imaging.

2. **Results and Discussion**

The LN powder consisting of microcubes with the size of 2-5 µm was obtained via a solvothermal process, starting from niobium pentoxide ($Nb_2O_5$) and lithium hydroxide (LiOH) precursors (see Experimental section for more details). The samples were prepared by spin coating LN powders diluted in ethanol on a substrate with patterned coordinates. An example of a single LN microcube is shown in the scanning electron microscope (SEM) image in Figure 1a. We conducted the measurement from only LN microcubes that are at least 20 µm apart from each other on the substrate (as shown in Figure 1b) to ensure that only emission from a single cube is collected as the excitation laser is tightly focused down to 10 µm with a lens and collected with a high numerical aperture (NA) objective (Figure 1a).

To study the photon-pair generation process, we performed the second-order auto-correlation measurements on a Hanbury-Brown-Twiss setup (Supporting information, Figure S3), using a tunable continuous wave (CW) laser with a center wavelength of ~780 nm to generate photons at ~1560 nm via SPDC. Figure 2a is a SEM image of a microcube with the size of ~4 μm. Although the LN cube is of rhombohedral structure, we can approximate that the width, length, and height are almost equal [35]. The correlation measurement on a LN microcube, acquired for 5 hours at pump energy of 60mW, demonstrated a strong correlation peak at zero-delay time. The coincidences-to-accidentals ratio (CAR) $g^{(2)}(0)$ -1 exceeds 2, indicating simultaneous arrival of the two photons (Figure 2b).

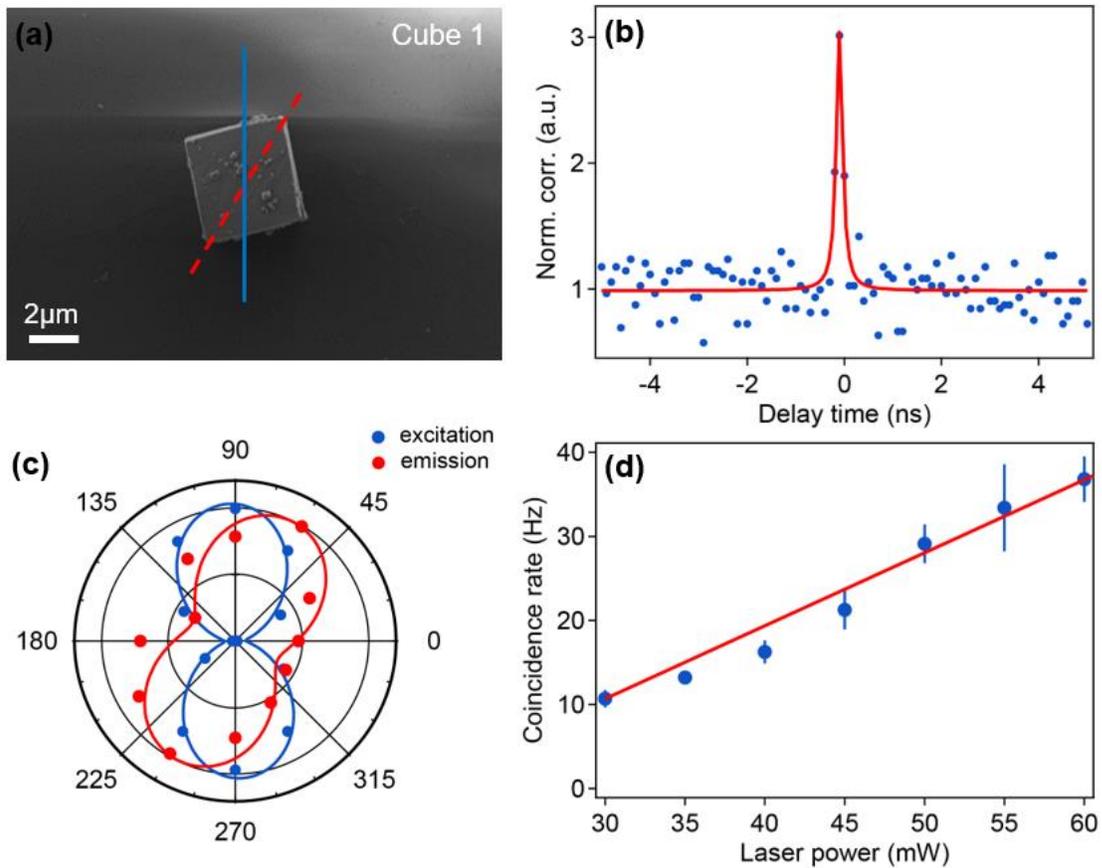

**Figure 2.** SPDC characterizations of a single microcube. (a) SEM image of a single LN microcube with the optical axis at 30° with respect to the polarization direction of input beam (90°). The optical axis (dashed red line) and the polarization axis of the laser (solid blue line) serve as a guide to the eye. (b) Autocorrelation function measurement on the cube in Figure 1a, demonstrates a correlation peak at zero-delay time. (c) Polarization plot of emitted photons from the cube (red circle), measured by extracting the correlation peak value at zero-delay time, intensity of the excitation laser without the sample (blue circle), and the fitting curves with $\cos^2\theta$ function. (d) Power dependence measurement of the coincidence rate as a function of excitation laser power with the linear fitting.

Figure 2c shows the polarization dependence of the emission from the cube (red circle) with respect to the intensity fluctuation due to polarization response of the fundamental pump laser (blue circle). All the intensity values in the radial axis are normalized from 0 to 1. The optical axis of the microcube is oriented along the diagonal of the rhombohedron, which is the projection of the polarization axis of the LN crystal on the lateral face [35]. Specifically, the red dashed line in Figure 2a is the direction of the optical axis, which yields an orientation of ~*30°* relative to the polarization direction of the pump (blue solid line). Thus, due to the form of $\chi^{(2)}$ susceptibility tensor of the material, SPDC emission will be maximized when the excitation laser is polarized along the optical axis of the cube, which is also shown in Figure 2b.

Next, we performed the power dependence measurements to further characterize the system. The coincidence rate was measured while changing the pump power from 30 mW to 60 mW (Figure 2d). The minimum excitation power is limited at 30 mW due to impractically long integration time for the correlation measurement at lower powers. The measurement demonstrates that the pair-generation rate increases linearly with the pump power [29]. After correcting for the transmission losses, the coincidence rate obtained with the specific cube (Figure 2a) is relatively modest at

~37 Hz, however, it is worth noting that the biphoton generation rate can suffer from scattering by crystal defects. We will discuss this point more in detail later in the text.

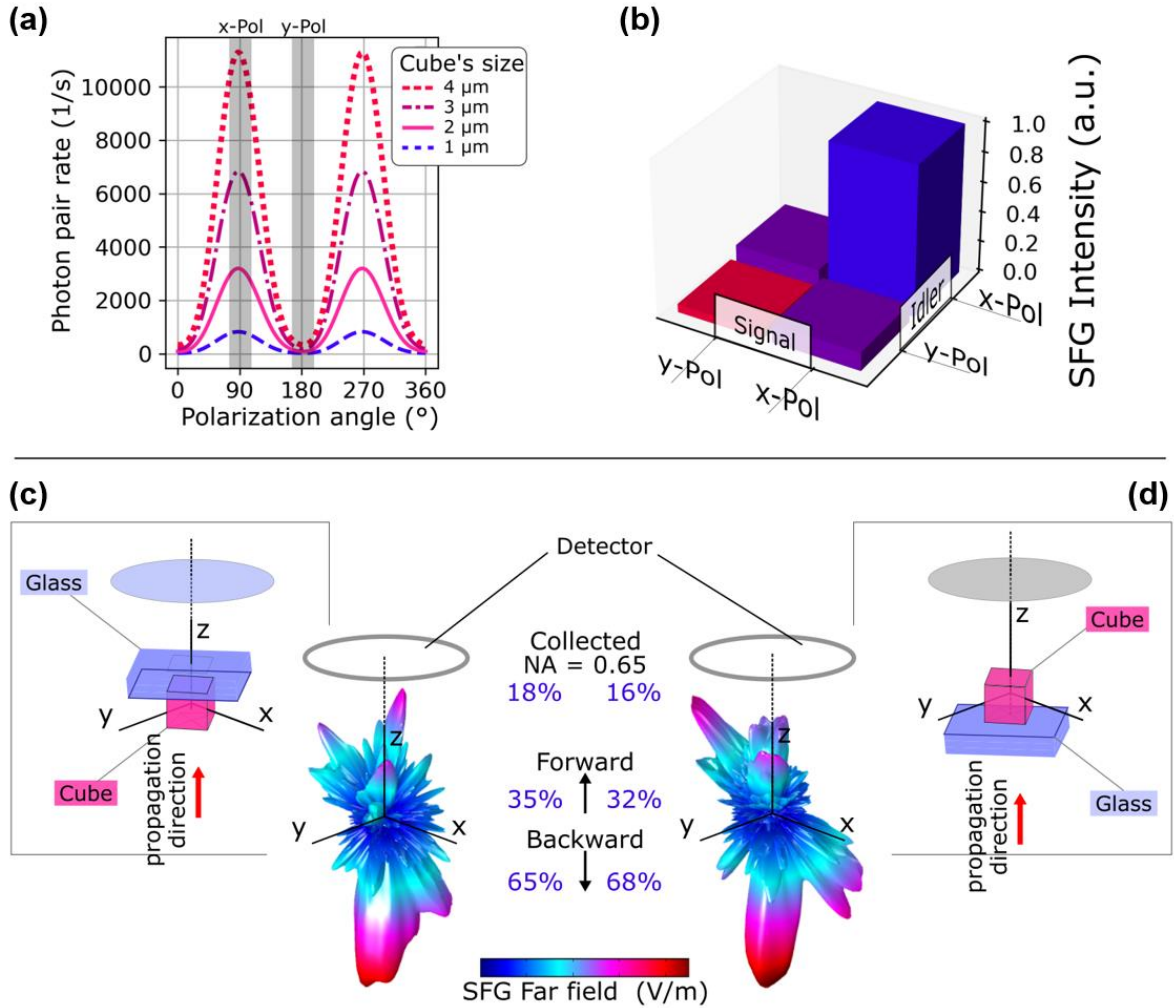

**Figure 3.** (a) Simulation results showing the input polarization dependence of the SPDC photon-pair rate generated from LN microcubes with different sizes. The biphoton rate is proportional to the size of the cube and maximal for the polarization angle of 90°/270°, considering the case where the beam is vertically polarized along the optical axis of the cube. (b) FEM simulation describing intensity of SFG emission depending on the linear polarization of the signal and idler photons. The intensity is calculated for a microcube with the size of 2 μm and is maximal for two photons linearly polarized along the optical axis of the microcube. (c-d) 3D schematics and far-field

radiation profile of the SFG signal from a microcube with the size of 2 μm for the two cases (c) the excitation wave propagates first through air, and the signal is collected on the glass side, and (d) the excitation wave propagates first through glass, and the SFG signal is collected on the air side. Comparatively more SFG is produced in forward direction as well as passes through a collection area (NA = 0.65) for the first case.

We estimated the two-photon generation rate from microcubes with different sizes by evaluating the efficiency of the inverse process of SPDC, namely sum-frequency generation (SFG), which has the same origin due to quantum classical correspondence principle [36][37]. The total photon pair generation rate across the area of the cube can then be calculated via the relation [37]:

$$\frac{1}{P_p}\frac{dN_{pair}}{dt} = 2\pi\eta^{SFG}\frac{\lambda_p^4}{\lambda_s^3\lambda_i^3}\frac{c\Delta\lambda}{\lambda_s^2} \qquad (1)$$

Here $P_p$ represents the pump beam power, and $dN_{pair}/dt$ is the photon-pair generation rate per unit signal frequency, and $\eta_{n_s n_i}^{SFG} = P_{SFG}/P_s P_i$ is the sum-frequency conversion efficiency, $\lambda_p$, $\lambda_s$, $\lambda_i$ are the pump, signal, and idler frequencies, respectively. In the degenerate SPDC case, $\lambda_s = \lambda_i = \lambda_p/2$, hence, Equation 1 becomes:

$$\frac{1}{P_p}\frac{dN_{pair}}{dt} = 2\pi\eta^{SHG}256\frac{c\Delta\lambda}{\lambda_p^4} \qquad (2)$$

with $\eta_{n_s}^{SHG}$ being the efficiency of the second-harmonic generation (SHG) process, which can be analytically calculated as described in the references [38]. The results were plotted in Figure 3a, demonstrating a ten-fold increase in the biphoton rate upon increasing the size of the cube from 1 to 4 μm. The predicted value for biphoton rate from a LN microcube of 4 μm is more than 10000 Hz, however, the experimental value can be lower than the theoretically predicted value, which may be attributed to the crystalline quality of the cube as well as the orientation of the optical axis.

To gain an insight into the emission of the biphotons, we performed a simulation using the Finite Element Method (FEM) (see Methods for more details) for the SFG process, and then predict the polarization entanglement and far-field emission pattern of the SPDC emission, which is proportional to the SFG amplitude of the classical signal and idler photons. In particular, we exploited the non-collinear detection scheme limited within the angle $\theta$ defined by the numerical aperture of the objective to predict the polarization states of generated photons. In this specific collection configuration, the generated photons are polarized in the x-y plane, hence, we explored the nonlinear process efficiency between x and y linear polarizations. Figure 3b presents the simulation results with a linear basis of signal and idler photons. The biphotons generated from the cube mostly share the same x-polarization due to the *d*-coefficients (Figure 3b).

Next, we predicted the radiation profiles of SPDC emission through SFG simulation. Specifically, we performed the simulation using the model where the cube was sitting on a glass substrate and excited from the glass side or the air side. Figure 3c demonstrates that biphotons are mainly generated in the forward or backward directivity along the z-direction, collinear with respect to the direction of the exciting pump wave vector. For LN, due to the C3v group of symmetry of the crystalline structure, the reduced nonlinear susceptibility tensor has the following structure:

$$d = \begin{pmatrix} 0 & 0 & 0 & 0 & d_{15} & -d_{22} \\ -d_{22} & d_{22} & 0 & d_{15} & 0 & 0 \\ d_{31} & d_{31} & d_{33} & 0 & 0 & 0 \end{pmatrix}$$

Here $d_{15} = d_{31} = -4.88 \frac{pm}{V}$, $d_{22} = 2.58 \frac{pm}{V}$ and $d_{33} = -34 \frac{pm}{V}$ [39]

This form of the $\chi^{(2)}$ tensor dictates the SPDC process through non-zero elements. Therefore, the orientation of the far field emission pattern along the c-axis can be attributed to the $d_{33}$ component, which has the largest value compared to the other *d*-coefficients. Furthermore, the nonlinear emission in opposite direction to the excitation propagation is due to the negative sign of this same component. The forward-to-backward ratios are 27% for the case of the glass above, i.e. the excitation laser coming directly from the side of the cube, meaning most emitted photons travel in a backward manner. When the cube was excited through the glass substrate, we obtained 37% forward emission. This configuration is particularly interesting as it addresses the problem of directive SPDC emission. The understanding of radiation patterns can help to further optimize the setup to achieve a higher biphoton collection efficiency. We also predict that SPDC emission from LN microcubes is more efficient when measured in reflection configuration.

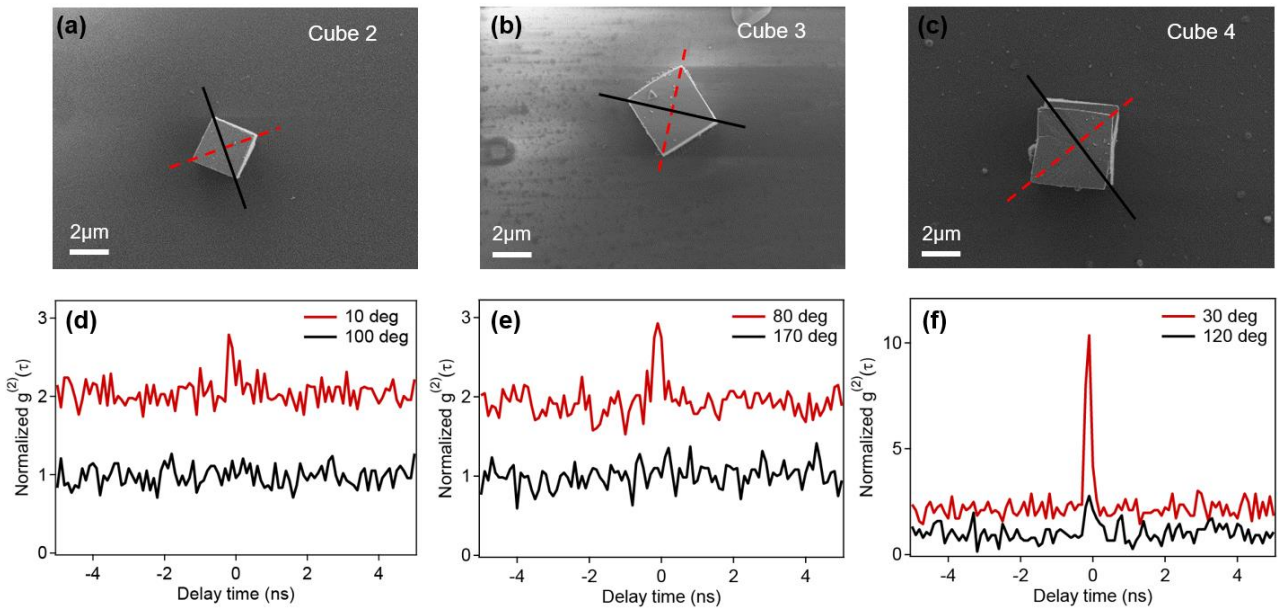

**Figure 4.** Polarization and size dependence of the emitted biphotons. (a-c) SEM images of the cubes 2, 3, and 4. The optical axis (dashed red line) and the polarization axis of the laser (solid blue line) serve as a guide to the eye. (d-f) Second-order autocorrelation function measurements from the cubes 2, 3, and 4 show the strongest correlation peak when the polarization of the excitation laser overlaps with the optical axis of the cube, whereas no correlation is obtained when the laser is retarded by 90°.

Finally, to showcase the agreement with the simulation studies, we performed the measurements on a different set of microcubes. Figures 4a-c present the measurements on three different cubes with the corresponding sizes of ~2.3, ~3.4, and ~4 µm respectively. The results of second-order autocorrelation function measurements obtained from three cubes are aligned with the calculated results, indicating the dependence of CAR on the size of the cubes. By increasing the size of the cubes, or increasing the interaction length, we obtained an improved CAR from 1.5 to 6 (Figure 4d-f), respectively corresponding to cube 2 (~2.3 µm) and cube 4 (~4 µm). Moreover, by optimizing the projection of pump polarization along the optical axis of the cube, we obtained the maximum rate of SPDC emission (Figure 4d-f). We also determined the SPDC conversion efficiency for all LN microcubes 1 to 4 by normalizing the biphoton count rate over the cube volume and pump power density. Those values are shown in Table 1. Specifically, we estimated that the biphoton rate (as shown in Table 1) increased ~2.1 times between cube 2 (~2.3 µm) and cube 1 (~4.1 µm), which is in good agreement with the predicted value of ~3.5 times (Figure 3a).

Notably, we obtained a SPDC conversion efficiency of ~1.2 GHz/Wm for cube 4 with acquisition time of 1 hour, which is significantly higher than the corresponding values for other cubes. There are several reasons attributed for this enhancement. Firstly, cube 4 has a larger dimension of ~ 4 µm compared to cube 2 (~2.3 µm) and cube 3 (~3.4 µm). Secondly, cube 4 possesses relatively pronounced and broad higher-order Mie resonances at both the pump and the decaying photon wavelengths (Supporting information Figure S1c). These resonances might improve the biphoton generation rate via the enhancement of the electric field within the resonant structure (Figure S2). Lastly, a clean surface with no apparent morphological defects might assist the nonlinear process in cube 4 (Figure 4c), whereas in cube 3, the presence of defects on the surface (Figure 4b) can dramatically suppress the nonlinear response [40].

**Table 1.** Summary of the nonlinear conversion efficiencies of 4 microcubes at 60 mW in this study.

| Cube number | Size [μm] | Biphoton count rate [Hz] | Conversion efficiency [GHz/Wm] |
|---|---|---|---|
| 1 | ~4.1 | ~37 | ~0.47 |
| 2 | ~2.3 | ~5.5 | ~0.22 |
| 3 | ~3.4 | ~6.6 | ~0.2 |
| 4 | ~4 | ~80 | ~1.2 |

## 3. Conclusions

In summary, we experimentally demonstrated biphoton generation via SPDC from a micro-sized LN cube. We obtained a photon-pair rate of 80 Hz from a ~4 μm size cube with the conversion efficiency of ~1.2 GHz/Wm, normalized to the excitation energy stored per unit volume. The photon-pair generation is supported by hybrid resonances of high order Mie and Fabry-Perot modes, surpassing the nonlinear efficiency of common bulky nonlinear crystals and on-chip devices. Using FEM modelling, we further presented the polarization correlation of the generated biphoton and their fair-field emission pattern. Our results demonstrate the capability of LN microcubes as an affordable and abundant nonclassical light source for flexible quantum-state engineering. Owing to the free-standing nature of the microcubes, they might be used as a versatile platform for integrated optical systems and free space communications. Also, due to the relation between the geometry of these cubes and crystal structure, the optical axes of these cubes are easily detectable. Therefore, by maximizing the projection of pump polarization along these axes, the maximum photon-pair generation rate can be achieved. Further investigation on different geometries of the cubes obtained via controlling various parameters during the growth process might also boost the SPDC efficiency. Finally, the successful measurement of two photon generation from a LN microcube via non-phase-matched SPDC opens up the potential for the development of quantum networking with different frequency channels, and broadband tunable squeezed states for sensing applications.

4. **Experimental Section**

*Fabrication of the sample:* $LiNbO_3$ particles are produced by a solvothermal synthesis method. Appropriate quantities of $Nb_2O_5$ (H.C. Starck, 99.92%) and LiOH (Aldrich, 98%) are dispersed in a mixture of Ethylene Glycol and distilled water. After a few minutes of ultrasonication, the suspension is poured into a PTFE coated stainless steel acid digestion bomb (model PA4748, volume 120 mL, Parr Instrument Company) and hydrothermally treated at 250° C for 70 h. After cooling, the reaction product is washed several times with water by centrifugation. Then, the final powder suspension is freeze dried. The size of the cubes can be varied by changing the Li / Nb molar ratio (R); in particular, with R=1.5 it is possible to obtain particles of 2-4 μm in size. LN powder was suspended in ethanol and spin coated on a substrate with alignment markers. After that, the sample was left to dry out at ambient temperature.

*Linear scattering cross-section measurement:* Linear scattering measurements were performed via dark field spectroscopy on a modified optical microscope. The light from the xenon arc lamp (150 W, Gilden Photonics) is focused onto the sample using a condenser (Zeiss 20× Epiplan-Neofluar dark field objective with an opaque disk to block the center of the light beam). The light scattered by the sample is collected with a 50× objective (Zeiss 50× Epiplan-Apochromat) and detected by a standard camera to check the microcube position. The signal was then coupled into a multimode fiber, where the fiber aperture of 200 μm serves as a confocal pinhole, which allows us to detect the signal from individual LNO cubes. The scattering spectra were obtained using a spectrometer (Andor Spectrometer) equipped with a visible and an IR CCD camera. All the linear scattering cross-section spectra were analyzed and background and dark noise corrected unless otherwise specified.

*SPDC emission probability*: SPDC studies were carried out using a home-built optical setup (Figure S3) with a tunable CW laser (Toptica) with a center wavelength of 780 nm. The excitation laser was focused on the sample by a lens (L1) with a focal length of 11 mm, and the emission was collected using a high numerical aperture (50×, NA = 0.65, Mitutoyo) objective lens. Two 1064 nm long-pass filters (LP) were used to block the residual pump. Afterwards, the obtained signal was directed to a fiber-based 50:50 beam splitter and detected by a superconducting nanowire single photon detector (SNSPD) (Single Quantum) with timing jitter of <10 ps. Correlation measurements were carried out using a time-correlated single-photon counting module (Swabian instruments).

*Simulation model:* We calculated the SPDC in Figure 3a with Equation (2) from the angle dependent power conversion efficiency provided by an analytical model which considers the length and the orientation of the nonlinear material with respect to the propagation direction of a specified pump beam [38]. The SFG in Figure 3b-c were calculated using a finite element method (COMSOL Multiphysics). The geometry for Figure 3b-c consisted in a spherical environment with the LN microcube in the middle. On one side of the microcube, the material was glass, while on the other, it was air. The field inside the microcube was calculated first for two independent linearly polarized light sources, for which the x and y polarization corresponded to the two projected diagonals of the cube. The polarization induced in the microcube, and which leads to SFG, was described with the LN $\chi^{(2)}$ tensor. The optic axis of LN was also considered and set as one of the microcube's diagonal. The $\chi^{(2)}$ tensor was rotated so that the optic/fast axis accordingly. The excitation was also x-polarization. Detailed simulation results are provided in the supporting information.

## Acknowledgements

We acknowledge the support of the Scientific Center for Optical and Electron Microscopy (ScopeM) of the Swiss Federal Institute of Technology (ETHZ). This work was supported by the Swiss National Science Foundation Grant 179099, the European Union's Horizon 2020 research and innovation program from the European Research Council under the Grant Agreement No. 714837 (Chi2-nano-oxides), Ambizione Grant No. 179966, and the Australian Research Council (grant number DE180100070).

Ngoc My Hanh Duong and Gregoire Saerens contributed equally to this work.

## Conflict of Interest

The authors declare no conflict of interest.